\documentclass[a4paper,twocolumn,10pt]{article}

\usepackage{siunitx}
\usepackage{tikz}

\usepackage{natbib}
\usepackage{amsmath}
\usepackage[font=small,labelfont=bf]{caption}
\usepackage{geometry}
\newgeometry{vmargin={15mm,25mm}, hmargin={15mm,15mm}}

\DeclareMathOperator{\dif}{d \!} 

\providecommand{\del}[2][-1]{
\ensuremath{\mathinner{
\ifthenelse{\equal{#1}{-1}}{ %
\left(#2\right)}{}
\ifthenelse{\equal{#1}{0}}{ %
(#2)}{}
\ifthenelse{\equal{#1}{1}}{ %
\bigl(#2\bigr)}{}
\ifthenelse{\equal{#1}{2}}{ %
\Bigl(#2\Bigr)}{}
\ifthenelse{\equal{#1}{3}}{ %
\biggl(#2\biggr)}{}
\ifthenelse{\equal{#1}{4}}{ %
\Biggl(#2\Biggr)}{}
}} %
}

\definecolor{mlbblue}  {rgb}{0.0000, 0.4470, 0.7410}
\definecolor{mlborange}{rgb}{0.8500, 0.3250, 0.0980}
\definecolor{mlbyellow}{rgb}{0.9290, 0.6940, 0.1250}
\definecolor{mlbpurple}{rgb}{0.4940, 0.1840, 0.5560}
\definecolor{mlbgreen} {rgb}{0.4660, 0.6740, 0.1880}
\definecolor{mlbcyan}  {rgb}{0.3010, 0.7450, 0.9330}
\definecolor{mlbred}   {rgb}{0.6350, 0.0780, 0.1840}
\definecolor{mlb1}{named}{mlbblue}  
\definecolor{mlb2}{named}{mlborange}
\definecolor{mlb3}{named}{mlbyellow}
\definecolor{mlb4}{named}{mlbpurple}
\definecolor{mlb5}{named}{mlbgreen}
\definecolor{mlb6}{named}{mlbcyan}
\definecolor{mlb7}{named}{mlbred}

\usepackage{ifthen}
\usepackage{soul}
\newboolean{markup}
\setboolean{markup}{false}
\ifmarkup
  \newcommand{\markup}[2]{\textsuperscript{\textbf{#1}}\hl{#2}}
\else
  \newcommand{\markup}[2]{#2}
\fi

\DeclareRobustCommand\sampleline[1]{
  \tikz\draw[#1] (0,0) (0,\the\dimexpr\fontdimen22\textfont2\relax)
  -- (2em,\the\dimexpr\fontdimen22\textfont2\relax);%
}

\begin{document}
\title{On the ice-nucleating potential of warm hydrometeors in mixed-phase clouds}
\author{
  Michael Krayer\textsuperscript{1}, 
  Agathe Chouippe\textsuperscript{1,a}, 
  Markus Uhlmann\textsuperscript{1}, 
  Jan Du\v{s}ek\textsuperscript{2}, 
  Thomas Leisner\textsuperscript{3}\\[0.5cm]
  \textsuperscript{1}\small{Institute for Hydromechanics, Karlsruhe Institute of Technology (KIT), Karlsruhe, Germany}\\
  \textsuperscript{2}\small{ICube, Fluid Mechanics Group, Université de Strasbourg, Strasbourg, France}\\
  \textsuperscript{3}\small{Institute of Meteorology and Climate Research, Atmospheric Aerosol Research Department,}\\\small{Karlsruhe Institute of Technology (KIT), Eggenstein-Leopoldshafen, Germany}\\
  \textsuperscript{a}\small{now at: ICube, Fluid Mechanics Group, Université de Strasbourg, Strasbourg, France}
  }
\date{}
\twocolumn[
  \begin{@twocolumnfalse}
    \maketitle
\begin{abstract}
The question whether or not the presence of warm hydrometeors in clouds may 
play a significant role in the nucleation of new ice particles has been debated
for several decades. While the early works of \cite{fukuta_numerical_1986} and
\cite{baker_nucleation_1991} indicated that it might be irrelevant, the more recent
study of \cite{prabhakaran_ice_2019} suggested otherwise. In this work, we are 
aiming to quantify the ice-nucleating potential using high-fidelity flow 
simulation techniques around a single hydrometeor and use favorable considerations
to upscale the effects to a collective of ice particles in clouds. While we find that ice nucleation
may be enhanced in the vicinity of a warm hydrometeor by several orders of magnitude
and that the affected volume of air is much larger than previously estimated, it is 
very unlikely that this effect alone causes the rapid
enhancement of ice nucleation observed in some types of clouds, mainly due to the low
initial volumetric ice concentration. Nonetheless, it is suggested to implement
this effect into existing cloud models in order to investigate second-order effects
such as ice nucleus preactivation or enhancement after the onset of glaciation.
\end{abstract}
  \vspace*{0.5cm}
  \end{@twocolumnfalse}
]

\section{Introduction}
\label{sec-intro}
The formation of hydrometeors in clouds is of great importance for the prediction of weather and cloud electrification, as well as
for the hydrological cycle, and thus, eventually for the evolution of climate. However, despite its relevance and great research effort over 
the past decades, many aspects remain poorly understood. One such puzzle is the discrepancy between the concentration of 
ice particles and that of available 
ice nuclei in airborne observations by several orders of magnitude \citep{pruppacher_microphysics_2010} which has been observed
for various cloud types
\citep{koenig_glaciating_1963,auer_observations_1969,hobbs_ice_1969,hobbs_ice_1985,mossop_origin_1985,hogan_properties_2002}.
This phenomenon has been termed \emph{ice enhancement} and various mechanisms which amplify primary ice nucleation 
have been proposed to explain the observed surplus in ice particles \citep{field_secondary_2016} . 

The most promising class of enhancement mechanisms is the so-called \textit{secondary ice production} (SIP), where new ice is formed
from preexisting ice particles. Commonly accepted SIP include rime-splintering \citep{hallett_production_1974}, fragmentation of ice
\citep{vardiman_generation_1978,takahashi_possible_1995,bacon_breakup_1998} and freezing drops \citep{hobbs_fragmentation_1968}.
Most of these mechanisms have been implemented into cloud models with explicit ice microphysics 
(see \cite{field_secondary_2016} for a recent overview), which, however, are not capable of satisfactorily explaining the
large amount of ice particles in observations. Further mechanisms have been proposed in the past 
whose relative importance is still to be evaluated. 

While studying the effect of supersaturation on primary ice formation, \cite{gagin_effect_1972} proposed several mechanisms 
which locally produce high values of supersaturation, and thus, regions of significantly enhanced nucleation activity. 
He suggested that freezing hydrometeors, which attain higher temperatures than the surrounding air due to the release of
latent heat, may cause transient supersaturation by simulateneous evaporation and heat transfer and linked it to the observation
of satellite drops which have previously been observed experimentally during the freezing of supercooled droplets by 
\cite{dye_influence_1968}. This hypothesis was corroborated by \cite{nix_nonsteady-state_1974}, who investigated numerically the 
transient freezing process of an isolated droplet. 

Another phenomenon which leads to localized supersaturation around hydrometeors is the riming process of ice particles in 
mixed-phase clouds with high liquid-water content \citep{gagin_effect_1972}.
Here, sufficiently large ice particles collect supercooled droplets from their surroundings, which then accumulate on its surface 
and subsequently freeze, leading to similar non-equilibrium conditions as observed for the freezing drop. 
The supersaturation around a riming ice particle has been investigated numerically by \cite{fukuta_numerical_1986} assuming 
steady conditions. Indeed, they found that the air in the vicinity of a warm ice particle can be highly supersaturated, while the 
magnitude and spatial extent strongly increase with increasing difference in surface to ambient temperature. However, even
at low cloud temperatures, the supersaturated regions do not extend far away from the hydrometeor in their simulations, which
led to the conclusion of \cite{rangno_ice_1991} that the overall ice enhancement is likely to be negligible.

While the previously mentioned studies focused on the quantification of the supersaturation field, \cite{baker_nucleation_1991} 
attempted to quantify the actual effect on ice enhancement around warm hydrometeors. He concluded that, although significant 
ice enhancement factors appear to be possible, the affected air mass seems to be too small to substantially contribute to
the explanation of the discrepancy between ice crystal and ice nucleus concentrations.

Recently, the potential of hail and rain to nucleate droplets has gained new attention when \cite{prabhakaran_can_2017}
observed condensation in the wake of cold droplets in a moist convection apparatus. In their experiment, a pressurized mixture 
of sulfur hexafluoride/helium was used and
operating conditions were chosen such that slight supersaturations led to homogeneous condensation, which could indeed be
observed in the wake of larger falling droplets. However, their results can only be qualitatively transfered to Earth's atmosphere
mainly due to the difference in primary nucleation mechanism. Moreover, the mechanism to create supersaturation is different
from the previously discussed one in the sense that the droplets are colder than the surrounding air.

To overcome these shortcomings, \cite{prabhakaran_ice_2019} conducted a similar laboratory experiment using moist air 
which has been seeded with aerosol particles as an operating medium to investigate heterogeneous nucleation in the 
wake of a warm falling droplet. Again, significant nucleation was observed in the wake of the falling hydrometeor. 
While the extent of wake-induced nucleation has not been quantified, 
the affected volume of air appears to be significantly larger than the predictions of \cite{fukuta_numerical_1986} and 
\cite{baker_nucleation_1991} were suggesting. This discrepancy might be related to the strong assumptions on the flow 
made in these early works, which neglect important features of the wake of falling objects that are nowadays 
accessible to numerical simulations \citep{johnson_flow_1999,bouchet_hydrodynamic_2006,zhou_chaotic_2015}.
In particular, it has been shown that both unsteadiness and vortical structures of the flow strongly affect the temperature and 
vapor concentration in the vicinity and far downstream of a falling sphere \citep{bagchi_direct_2001,de_stadler_large_2014},
and thus strongly affect the distribution of supersaturation around hydrometeors \citep{chouippe_heat_2019}.

In \cite{chouippe_heat_2019}, we presented a framework for high-fidelity numerical simulations of heat and
mass transfer around a falling ice particle which is not in thermal equilibrium with its surroundings. 
Even though the focus was predominantly set towards several methodological questions, it was shown qualitatively that 
the local supersaturation differs strongly from the simpler considerations of \cite{fukuta_numerical_1986}.

The present work aims to revisit the details of the supersaturation field around an idealized falling hydrometeor and
to quantify the volume of air which is affected by the presence of hydrometeors.
Furthermore, it is yet to be evaluated how significant meteor-induced enhancement of ice nucleation is in clouds
\citep{korolev_new_2020}, and therefore, we attempt to link our results to heterogeneous nucleation of aerosol particles.

\section{Methodology}
\label{sec:method}
In order to assess the spatial structure of supersaturation around hydrometeors,
knowledge on the flow around it is essential. This is not an easy task, since
the features of the flow may strongly depend on parameters such as the size, shape
and surface properties of the hydrometeor, which typically vary substantially, especially if
ice particles are concerned. %

In this work we utilize the numerical framework previously presented in \cite{chouippe_heat_2019}
and model the hydrometeor as a sphere with constant diameter $D$, constant surface temperature $T_p$ and 
a constant vapor pressure $e_{v,p}$ at its surface. The latter is determined by the assumption of local
equilibrium at the surface, and thus, corresponds to the saturation vapor pressure $e_{sat,j}$ at $T_p$.
Here, $j$ denotes the phase of water on the surface of the hydrometeor, where subscript "$i$" will be used to
denote the solid phase and "$w$" to denote the liquid phase.

The boundary conditions for the heat and mass transfer problem are motivated by the conditions expected 
for riming ice particles in the wet-growth regime. A surface temperature of $T_p = 0\,\si{\celsius}$ 
is assumed, i.e. the surface of the ice sphere is warmer than the local cloud temperature, since the latter is
generally below the freezing point in mixed-phase clouds. Please note that at this surface temperature
the equilibrium vapor pressure with respect to both the liquid and solid phase coincide, and hence, no assumptions on the
phase at the surface have to be made.
Since the actual riming process is not modelled explicitly,
but only enters through the boundary conditions of the heat and mass transfer problem, the results of this work 
are more general and can in principle be applied to any configuration which contains a mechanism leading to a warm hydrometeor,
such as a freezing droplet or rapid decreases in cloud temperature.

One of the key parameters for the investigation of wake-induced ice nucleation is the distribution of the
saturation ratio 
\begin{equation}
S_j(e,T) = e/e_{sat,j}(T),
\end{equation}
where $e$ denotes the local vapor pressure. A parametrization of $e_{sat,j}$ with respect to
temperature for both phase equilibria in the temperature range of interest is adopted from \cite{murphy_review_2005}. 
The spatial distribution of $S_j$ can then be reconstructed from the temperature field $T$
and the vapor concentration $n_v$, which are governed by the transport equations
\begin{eqnarray}
   \frac{\partial T}{\partial t} + \vec{u}\cdot \vec{\nabla} T
      &=&
      \mathcal{D}_T\vec{\nabla}^2 T, \label{eq:NS_heat}\\
   \frac{\partial n_v}{\partial t} + \vec{u}\cdot \vec{\nabla} n_v
      &=&
      \mathcal{D}_{n_v}\vec{\nabla}^2 n_v \label{eq:NS_mass},
\end{eqnarray}
where $\mathcal{D}_T$ and $\mathcal{D}_{n_v}$ denote the heat and vapor diffusivities and $\vec{u}=(u,v,w)^T$ is the
velocity field of the surrounding air. The partial pressure of water is linked to these transported quantities by 
\begin{equation}
e = n_v / k_b T,
\end{equation}
where $k_b$ is Boltzmann's constant. Even though the diffusivities of both vapor and heat are similar in magnitude,
it was shown by \cite{chouippe_heat_2019} that assuming equal diffusivities leads to an underestimation of the
saturation ratio, and hence, both equations will be treated separately.

Under the boundary conditions described earlier, Eq.~\eqref{eq:NS_heat} leads to an outward heat
flux, because the surface temperature of the hydrometeor is higher than that of its surrounding. Furthermore,
at $T_p = 0\,\si{\celsius}$ the vapor concentration at the meteor's surface
is generally higher than in the ambient, since latter is assumed to be in equilibrium with the liquid phase
at the corresponding cloud temperature $T_\infty < T_p$ due to the presence of
droplets in mixed-phase clouds, i.e. $e_\infty = e_{sat,w}(T_{\infty})$. This results in an outward vapor flux, and hence,
the hydrometeor will also evaporate. It is therefore of primary importance to resolve the temporal and spatial variations
of both temperature and concentration fields.%

In \cite{chouippe_heat_2019} we have shown that buoyant forces due to density variations within
the fluid phase caused by variations in temperature and concentration are negligible at the parameter point 
of interest. The flow can then be approximated by the incompressible Navier-Stokes equations,
\begin{eqnarray}
   \vec{\nabla} \cdot {\vec{u}} 
      &=& 0, \label{eq:NS_continuity}\\
   \frac{\partial \vec{u}}{\partial t} + (\vec{u}\cdot \vec{\nabla}) \vec{u}
      &=&
      -\frac{1}{\rho_f}\vec{\nabla} p + \nu \vec{\nabla}^2 \vec{u}, \label{eq:NS_momentum}
\end{eqnarray}
where $p$ denotes the hydrodynamic pressure, $\rho_f$ the density and $\nu$ the kinematic viscosity of the fluid.
The flow is driven by the falling motion of the hydrometer, which is assumed to be moving
with a constant velocity $\vec{v}_p$ equal to its terminal velocity $v_T$, i.e.
\begin{equation}
\vec{v}_p = -v_T \vec{e}_z,
\end{equation}
where $\vec{e}_z$ is the unit vector in the $z$-direction. The fluid ahead of the hydrometeor is assumed to be at rest.
This setup is equivalent to a system with a fixed sphere positioned in an 
upcoming flow, which differs from that of a freely falling mobile sphere in the sense that fluctuations 
in $\vec{v}_p$ are not permitted. The necessity to account for these variations has been discussed in \cite{chouippe_heat_2019} 
and it was found that it has little influence in the present context. 

\begin{figure}[t!]
  \centering
  \includegraphics[]{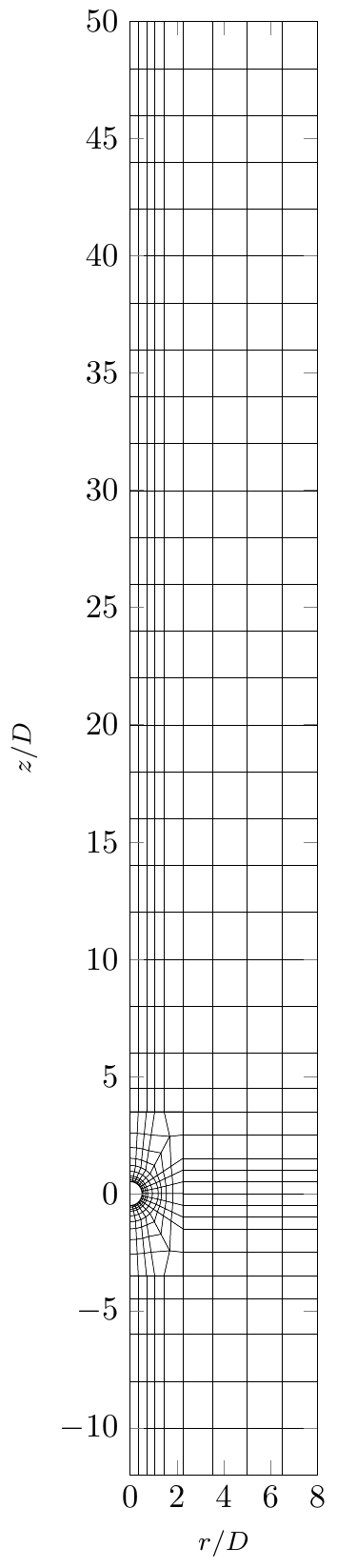}
  \caption{
      Spectral-element mesh in the axial-radial plane, where the lines depict the boundaries of the $463$
      elements. Each element contains ${[}6 \times 6 {]}$ collocation points. The three-dimensionality is introduced
      by Fourier expansion in the azimuthal direction, which is truncated at the 7th Fourier mode, 
      \markup{2.4}{resulting in a cylindrical domain. A uniform velocity profile with constant temperature and vapor content is
      imposed at the upstream boundary, while the downstream boundary is subject to zero-gradient BCs for both velocity and scalars.
      The lateral boundaries are stress-free at zero pressure with zero-gradient BCs for the scalars.
      The surface of the spherical particle, whose center is located at the origin of the coordinate system, is impermeable and a no-slip BC 
      for velocity is imposed as well as constant Dirichlet BCs for the scalar fields.}
    }
  \protect\label{fig:semmesh}
\end{figure}

The Navier-Stokes and scalar transport equations are solved numerically in non-dimensional form.
\markup{2.4}{
In particular lengths are scaled by the particle diameter $D$, velocity components by the terminal velocity $v_T$ and the scalar
transport equations are formulated in terms of}
\begin{align}
\tilde{T}   &= (T - T_{\infty}) / (T_p - T_{\infty}), \\
\tilde{n}_v &= (n_v - n_{v,\infty}) / (n_{v,p} - n_{v,\infty}),
\end{align}
\markup{}{which has the advantage, that various
dimensional boundary conditions, e.g. various values of $T_{\infty}$, can be studied in post-processing from a single simulation run.}
Under the simplifications stated, the non-dimensional problem can be parametrized by the Reynolds number, $Re = \left| \vec{v}_p \right| D / \nu$, the
Prandtl number, $Pr = \nu / \mathcal{D}_T$, and the Schmidt number, $Sc = \nu / \mathcal{D}_{n_v}$. While the Reynolds number is varied 
up to a value of $Re = 600$, the Prandtl and Schmidt numbers 
are set, respectively, to values of $Pr = 0.72$ and $Sc = 0.63$ in this study, which correspond to representative values expected for humid air
in the temperature range of interest.

The numerical method employed is based on the method of \cite{jenny_efficient_2004}, which has been used by e.g. \cite{kotouc_loss_2008} 
and \cite{chouippe_heat_2019} to simulate heat and mass transfer around spherical particles.
It relies on a spectral/spectral-element discretiztion of the Navier-Stokes equations (Eq. \ref{eq:NS_continuity},\ref{eq:NS_momentum}) 
coupled to the transport of heat and mass (Eq. \ref{eq:NS_heat},\ref{eq:NS_mass}). 
The spherical particle is placed at the origin of a cylindrical domain $\Omega$.
\markup{2.5}{
On the surface of the particle impermeability and no-slip boundary conditions are imposed, while the scalar fields are subject to
the constant Dirichlet boundary conditions $\tilde{T} = \tilde{n}_v = 1$. At the upstream boundary of the simulation domain, 
a uniform velocity condition and $\tilde{T} = \tilde{n}_v = 0$ are enforced, while the downstream boundary is subject to zero-gradient
boundary conditions for velocity as well as the scalars. The lateral boundaries are stress-free with zero-gradient boundary conditions
for the scalars and zero pressure is imposed.
The axial-radial plane is decomposed using the spectral element method of \mbox{\cite{patera_spectral_1984}}, while the homogeneous
azimuthal direction is treated using Fourier decomposition. The numerical mesh used in this work is shown in \mbox{fig.~\ref{fig:semmesh}}.}
\markup{2.6}{
For the temporal integration a time splitting method is used which consists of an explicit third order Adams-Bashforth
discretization for the advective terms and a first order fully implicit discretization of the diffusion terms \mbox{\citep{ronquist_optimal_1988}}.
}

In the vertical direction the domain has a total length of $62 D$, where the inflow length is $12 D$ and the outflow length $50 D$
measured from the center of the sphere. The diameter of the domain is $16 D$. 
The computational domain has therefore been extended in the rear of the sphere compared to our former simulations 
presented in \cite{chouippe_heat_2019}.
In total, 463 two-dimensional elements, each containing ${[}6 \times 6 {]}$ collocation points, have been distributed over the domain. 
The azimuthal Fourier series is truncated at the 7th mode. This resolution is comparable to our previous work and has been shown to give good results
for the momentum, as well has heat and mass transfer. All relevant scales of the flow and the scalar fields are resolved.

For more details on the numerical framework, the reader is referred to the previous work of \cite{chouippe_heat_2019}.

\section{Results}
\label{sec:results}

\subsection{Hydrometeor wake regimes in clouds}
\label{sec:wake-regimes}
The dynamics of the flow around spherical objects are, under the assumptions stated in the previous section, fully
parametrized by the Reynolds number, which depends on the diameter and indirectly, through modulation of the terminal velocity, 
on the density of the hydrometeor. Depending on the Reynolds number value, various flow states emerge in the 
wake \citep{johnson_flow_1999,jenny_instabilities_2004,kotouc_transition_2009}.
\markup{2.1}{At low values the wake is axisymmetric and steady. When the first critical value of the Reynolds number, $Re_{c,1} \approx 212$, is exceeded,
the wake becomes oblique with respect to the falling direction (\emph{steady oblique regime)} and only planar symmetry may be observed \mbox{\citep{ghidersa_breaking_2000}}. 
At $Re_{c,2} \approx 273$ a second instability of Hopf type occurs \mbox{\citep{ghidersa_breaking_2000}}, which leads to unsteady periodic vortex shedding, 
while planar symmetry is still maintained (\emph{oscillating oblique regime}). Eventually the vortex shedding becomes chaotic at 
$Re_{c,3} \approx 360$ \mbox{\citep{ormieres_transition_1999}} and all instantaneous symmetries are lost (\emph{chaotic regime}). 
The listed regimes differ significantly in their ability to transfer heat and mass \mbox{\citep{chouippe_heat_2019}}, and thus, different characteristics
in producing local supersaturation are to be expected.}

\markup{}{In order to estimate the distribution of ice particle wake regimes in clouds}, we adopt the size 
distribution of \cite{marshall_distribution_1948},
\begin{equation}
  N = N_0 \exp \del{-\lambda D},
  \label{eq:marshall_palmer}
\end{equation}
where $N$ is the number concentration density of hydrometeors per unit volume of air and $N_0, \lambda$ are model parameters.
Equation~\eqref{eq:marshall_palmer} has originally been developed for raindrops, however, its validity for sufficiently large 
ice particles has been 
demonstrated by \cite{passarelli_theoretical_1978,houze_size_1979,gordon_airborne_1984,herzegh_size_1985,patade_particle_2015}
for various types of natural clouds. 
The values of the model parameters $N_0,\,\lambda$ are usually obtained by airborne measurements and given as 
functions of the cloud temperature. 
\markup{2.2}{
One such parametrization is provided by \mbox{\cite{houze_size_1979}} for frontal clouds 
within a temperature range from $-42\si{\celsius}$ to $+6\si{\celsius}$, which will
be used in the following for the size distribution of primary ice and is given by
the constitutive equations}
\begin{align}
 N_0 (T)     &= 5.5 \cdot 10^6 \exp\left( -0.09 T \right) \si{\per\meter\tothe{4}}, \\
 \lambda (T) &= 9.6 \cdot 10^{-1} \exp\left( -0.056 T \right) \si{\per\milli\meter},
\end{align}
\markup{}{where $T$ denotes the cloud temperature in $\si{\celsius}$. For this cloud type, ice enhancement has been 
previously reported in the presence of riming ice particles \mbox{\citep{hogan_properties_2002}}.}
Even though it is known that the number density of ice particles of sub-millimeter size might deviate significantly 
from Eq.~\eqref{eq:marshall_palmer} in a sub-exponential or super-exponential manner \citep{passarelli_theoretical_1978}, 
depending on cloud conditions, we assume that Eq.~\eqref{eq:marshall_palmer} holds nonetheless for all ice particle sizes 
for the sake of simplicity.

For reasons explained in section~\ref{sec:global-ice-enhancement}, our quantity of interest is not the number concentration, 
but rather the volume fraction $\phi$ of hydrometeors in clouds, which can be obtained 
by integrating the corresponding moment of the Marshall-Palmer distribution, i.e.
\begin{equation}
  \phi = \int_{0}^{\infty} \frac{\pi D^3}{6} N_0 \exp \del{-\lambda D} \dif D.
\end{equation}
\markup{2.3}{In order to determine the wake regimes from the size distribution, the terminal velocity needs to be approximated as a function
of the ice particle diameter, and hence, further assumptions on the density of the ice particles have to be made.}
For graupel particles, the density may take values which range between 
$0.05\,\si{\gram\per\cubic\centi\meter}$ and $0.89\,\si{\gram\per\cubic\centi\meter}$, depending on growth conditions and history 
\citep{pruppacher_microphysics_2010}. For our estimation we choose a value of $0.6\,\si{\gram\per\cubic\centi\meter}$, as we are primarily concerned with 
ice particles in the wet-growth mode, which typically constitute the upper end of the density range. 
\markup{2.3}{
Using this value and the empirical drag law of \mbox{\cite{schiller_uber_1933}}, the critical diameters for regime transition are calculated
from the critical Reynolds numbers.
The total solid volume fraction can then be subdivided
into the contributions by ice particles with a certain wake regime, i.e.}
\begin{equation}
  \phi \equiv \sum_j \phi_j = \sum_{j} \int_{D_{j,\min}}^{D_{j,\max}} \frac{\pi D^3}{6} N_0 \exp \del{-\lambda D} \dif D,
  \label{eq:phij}
\end{equation}
where the index $j$ denotes one of the four wake regimes (steady axisymmetric; steady oblique; oscillating oblique; chaotic) and $D_{j,\min}$, 
$D_{j,\max}$ the corresponding lower/upper limit for the hydrometeor diameter to be in this regime. 

\begin{figure}[t!]
  \centering
  \includegraphics{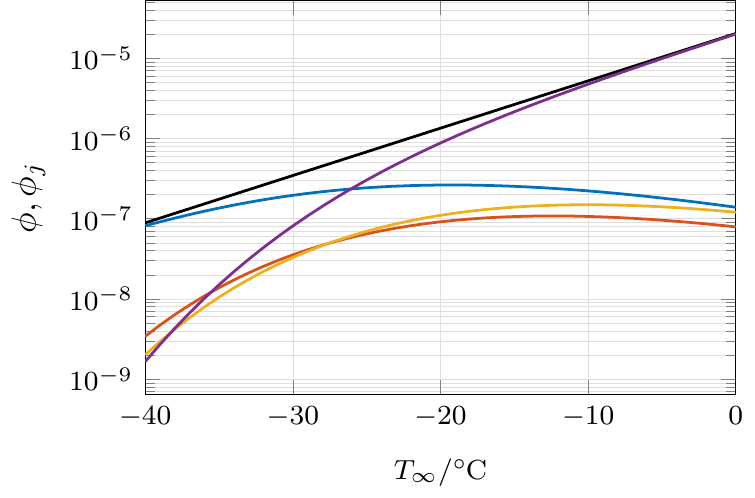}
  \caption{
      Volume fraction of ice particles in clouds as a function of ambient temperature. The total volume fraction
      $\phi$ is displayed by a solid black line ($\sampleline{black}$), while the contribution by the regimes is
      given by the colored lines.
      Linestyles: axisymmetric regime ($\sampleline{mlb1}$), steady oblique regime ($\sampleline{mlb2}$),
      oscillating oblique regime ($\sampleline{mlb3}$), chaotic regime ($\sampleline{mlb4}$).
    }
  \protect\label{fig:volume_fraction}
\end{figure}

\begin{figure}[th]
  \centering
  \includegraphics{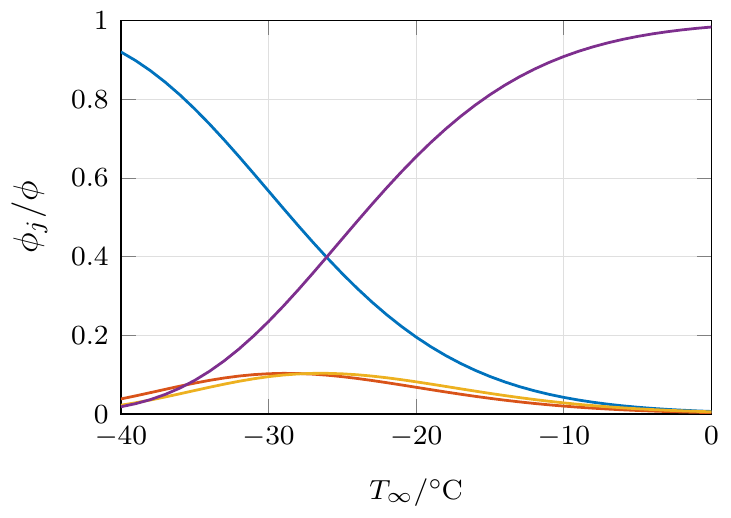}
  \caption{
      Relative contribution to the total volume fraction by hydrometeors in a certain regime.
      Linestyles: axisymmetric regime ($\sampleline{mlb1}$), steady oblique regime ($\sampleline{mlb2}$),
      oscillating oblique regime ($\sampleline{mlb3}$), chaotic regime ($\sampleline{mlb4}$).
    }
  \protect\label{fig:regime_contribution}
\end{figure}

Figure~\ref{fig:volume_fraction} shows the estimated volume fraction of frozen hydrometeors in clouds as a function
of cloud temperature. The values are typically smaller than $10^{-5}$ and decrease exponentially with decreasing
temperature. When comparing the various regimes, it can be observed that the largest contributions generally stem
from hydrometeors in the axisymmetric or chaotic regime. This becomes especially clear when looking at the relative
contribution of $\phi_j$ to the total volume fraction $\phi$ as shown in fig.~\ref{fig:regime_contribution}. As can be
seen, more than $80\si{\percent}$ of ice particles (by volume) exhibit one of these two wake regimes. Chaotic wakes
are dominant at temperatures close to the freezing point, while axisymmetric wakes dominate when the temperature is very
low. We will therefore focus solely on the axisymmetric and chaotic regimes in the following.

\subsection{Supersaturation in the wake of hydrometeors}
\begin{figure}[th]
  \centering
  \includegraphics[]{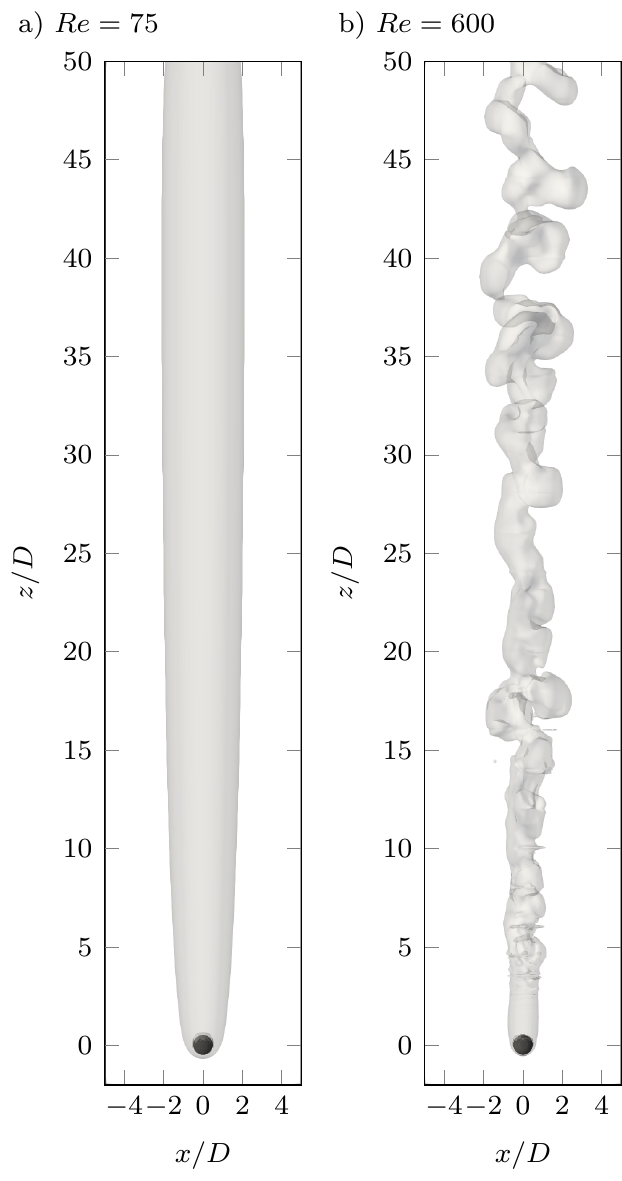}
  \caption{
      Isosurfaces of supersaturation in the wake at $T_{\infty} = -30 \si{\celsius}$. The
      value of the isocontour is $\tilde{s}^*_i = 0.1$, i.e. ten percentage points higher than the
      ambient supersaturation. Two different wake regimes are depicted, which correspond to two different
      hydrometeor sizes in our framework. (a) axisymmetric regime at $Re = 75$, (b) chaotic regime at $Re = 600$.
    }
  \protect\label{fig:supsat_instant}
\end{figure}

We now examine the saturation profiles in the wake of hydrometeors in the two regimes of interest. 
Figure~\ref{fig:supsat_instant} shows isocontours of supersaturation w.r.t. ice, defined as $s_i = S_i - 1$, for two different 
values of the Reynolds number, namely $Re = 75$, which lies within the axisymmetric regime, and
$Re = 600$ in the chaotic wake regime. As ambient temperature, a value of $T_{\infty} = -30\si{\celsius}$
was adopted, because this roughly corresponds to the cloud temperature where the two regimes are of equal
significance (cf. fig.~\ref{fig:regime_contribution}). Since the ambient fluid is already supersaturated 
w.r.t. ice, the threshold of the isocontours is given as an excess to the value in the ambient and we
define the excess supersaturation as
\begin{equation}
  \tilde{s} \equiv s - s_\infty.
\end{equation}

\begin{figure}[th]
  \centering
  \includegraphics[]{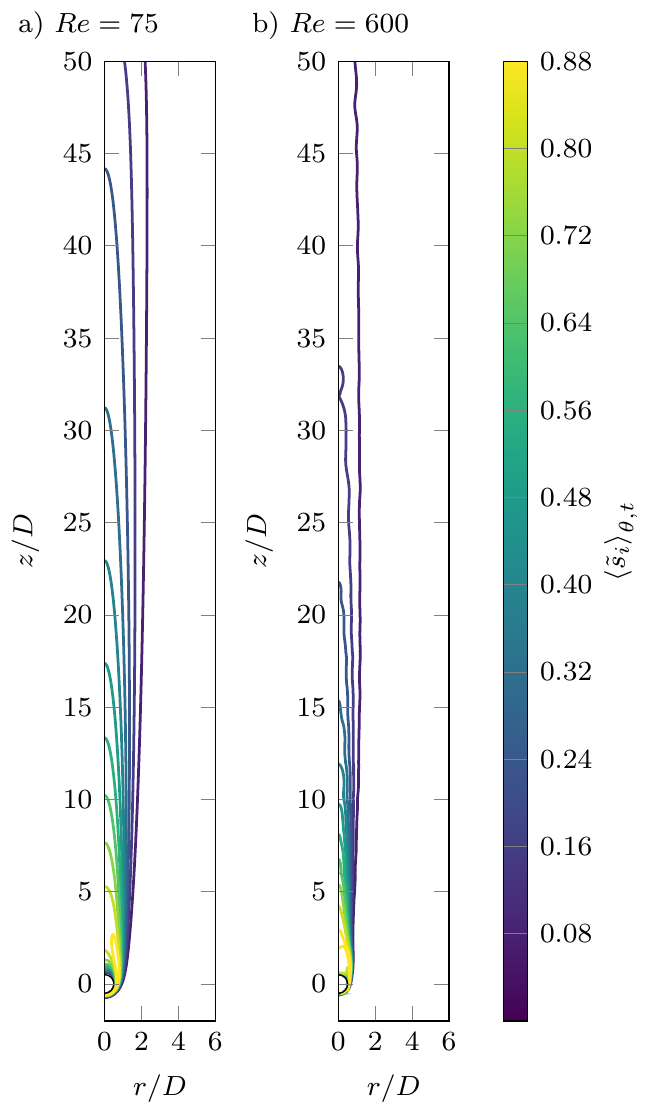}
  \caption{
      Contours of excess supersaturation in the wake, averaged over time and azimuthal direction at $T_{\infty} = -30 \si{\celsius}$.
      (a) axisymmetric regime at $Re = 75$, (b) chaotic regime at $Re = 600$.
    }
  \protect\label{fig:supsat_average}
\end{figure}

At $Re = 75$, the flow is steady, and thus, so is the supersaturation field. In the chaotic regime, 
the flow is characterized by time-dependent vortex shedding from the ice particle's boundary layer, and therefore 
excess supersaturation appears intermittently. In both regimes, significant excess supersaturation w.r.t. the 
ambient can be observed far downstream and the volume of air which is affected by the wake is by far larger 
than the volume of the meteor. This can be seen more clearly when averaging the fields in the azimuthal direction,
which is statistically homogeneous for both cases, as well as in time for the unsteady flow. The averaged
excess supersaturation w.r.t. ice is shown in fig.~\ref{fig:supsat_average} for both cases. 
The saturation profiles differ substantially from the ones obtained by \cite{fukuta_numerical_1986} for similar 
boundary conditions and Reynolds number. This is presumably due to their strong simplification of potential flow,
which is incapable of reproducing the boundary layer and the flow in the wake of the ice particle correctly. 
This indeed leads to strong modifications in the distribution
of supersaturated regions. Regions of high vapor content and relatively low temperatures, i.e. supersaturated regions,
which are created in the mixing layer close to the riming particle are transported further downstream by the detaching 
vortices in the chaotic regime. This effect does not occur in the simulations of \cite{fukuta_numerical_1986}, who observed 
a complete decay of
supersaturation to the inflow value within a distance of approximately $2.5 D$ downstream of the ice particle for 
$T_{\infty} = -30 \si{\celsius}$ 
while fig.~\ref{fig:supsat_average} clearly indicates that regions with $\tilde{s}_i > 0$ may
be observed at distances more than $50 D$ from the rear at this temperature.

\begin{figure}[th]
  \centering
  \includegraphics[]{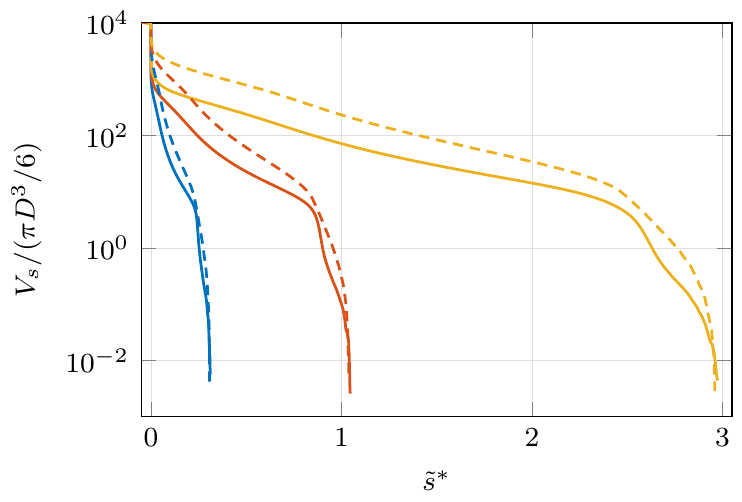}
  \caption{
      Volume of air where supersaturation exceeds a given threshold as a function of the
      threshold. The volume is normalized by the volume of the ice particle and three different temperatures
      are shown: 
      $T_{\infty} = -20\si{\celsius}$ ($\sampleline{mlb1}$),
      $T_{\infty} = -30\si{\celsius}$ ($\sampleline{mlb2}$),
      $T_{\infty} = -40\si{\celsius}$ ($\sampleline{mlb3}$).
      Solid lines correspond to $Re = 600$ (chaotic regime), while dashed lines show the data obtained for
      $Re = 75$ (axisymmetric regime).
  }
  \protect\label{fig:supsat_volume_cdf}
\end{figure}

In order to compare the induced supersaturation for different values of $T_{\infty}$ and in between different regimes, we compute 
the time-averaged volume of air which is supersaturated above a given threshold,
\begin{equation}
  V_s(s^*) \equiv \frac{1}{\tau}\int_{0}^{\tau}\int_{\Omega} H(s(\vec{x},t)-s^*) \dif \Omega \dif t,
  \label{eq:cdf-supsat-vol}
\end{equation}
where $V_s$ is the superaturated volume, $s^*$ the corresponding threshold and $H$ the Heaviside step function. 
Figure~\ref{fig:supsat_volume_cdf} shows this quantity normalized by the volume of the ice particle as a function of the threshold. 
It can be seen, that in the axisymmetric regime, the supersaturated volume is generally larger than in the
chaotic regime. This is caused by the enhanced mixing properties of the vortical wake structures in the latter case,
which lead to a faster decay of the scalar field. 

For both regimes, the range of observed values of supersaturation is similar.
In fact, differences in range may only occur due to the distinct diffusivities of the temperature and water vapor fields, which 
indeed lead to subsaturations in the chaotic regime in some parts of the flow.
While the range can be estimated reasonably well using simple mixing arguments \citep{chouippe_heat_2019}, we observe that
the highest values of supersaturation only occur in a very confined portion of the wake. 
\markup{2.7}{However, if the temperature difference 
between the ambient and the ice particle is sufficiently high, a strong excess in supersaturation compared to the ambient
value may occur in a volume comparable to the size of the meteor, if no mechanism for depletion is 
considered. This excess may be as high as 270\% at a temperature difference of $40\si{\kelvin}$.}
For all values of $T_{\infty}$ the
volume of air which is affected by the wake-induced supersaturation is of the order of $10^3$ particle volumes.

\subsection{Ice enhancement due to meteor wakes}
\label{sec:global-ice-enhancement}
Our spatially resolved data allows us, for the first time to our knowledge, to quantify the significance of
wake-induced supersaturation on ice enhancement. Thus, in this section, we are now aiming to estimate how the nucleation
of ice is affected by this excess in supersaturation. We therefore adopt the commonly used ice enhancement factor 
\citep[e.g.][]{baker_nucleation_1991}, which
is defined as the ratio between the observed number concentration of ice nuclei (IN) and that expected for a reference state.
For our problem, it will be defined as
\begin{equation}
  f_i \equiv \frac{N_{IN} \,\text{(with meteors)}}{N_{IN} \,\text{(without meteors)}}.
\end{equation}
The IN number concentration $N_{IN}$, i.e. those aerosol particles which can be activated at a given supersaturation, is 
commonly approximated by a power-law in terms of the supersaturation w.r.t. ice,
\begin{equation}
  N_{IN} = C s_i^{\alpha}
  \label{eq:huffman}
\end{equation}
where $C, \alpha$ are constants depending on the aerosol composition \citep{huffman_supersaturation_1973}. For natural aerosols, the exponent
$\alpha$ typically ranges between $3$ and $8$, where larger values are attributed to more polluted air masses, e.g. near metropolitan or
industrial areas \citep{pruppacher_microphysics_2010}. 
We are aware that Eq.~\eqref{eq:huffman} is a very simplified approach to the problem and its validility might be debatable. In particular, it does
not take into account the various characteristics of the different freezing modes, which are dependent on the temperature and supersaturation
w.r.t. the liquid phase, nor the time scales of freezing, which might play an significant role in the microphysical view. The results should 
therefore only be regared as a first crude estimation on the importance of wake-induced ice enhancement in clouds.

\begin{figure}[th]
  \centering
  \includegraphics[]{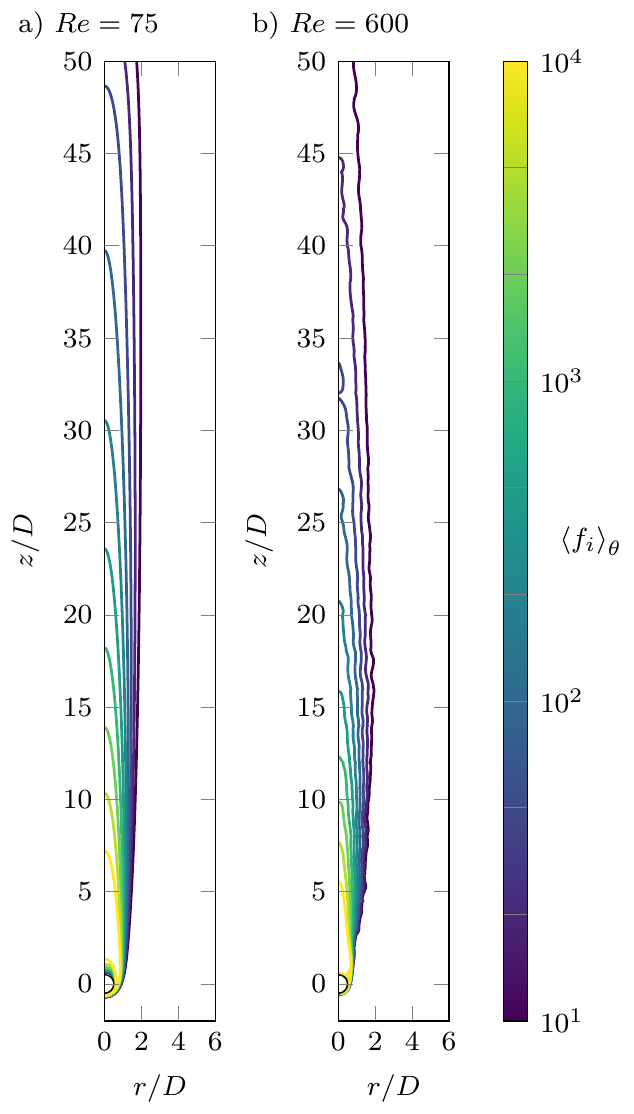}
  \caption{
      Contours of local ice enhancement factor in the wake, averaged over time and azimuthal direction at $T_{\infty} = -30 \si{\celsius}$.
      The contour lines are spaced logarithmically.
      (a) axisymmetric regime at $Re = 75$, (b) chaotic regime at $Re = 600$.
    }
  \protect\label{fig:contour_enhancement_Tinf243_alpha8}
\end{figure}

Using Eq.~\eqref{eq:huffman}, the local ice enhancement factor can be expressed as
\begin{equation}
  f_i(\vec{x}) = \frac{\left<s_i^{\alpha}\right>_t(\vec{x})}{s_{i,\infty}^{\alpha}}.
  \label{eq:fix}
\end{equation}
This quantity gives a quantitative measure on how much more ice is produced in the wake of the hydrometeor compared to the ambient 
under the limitations stated above. Figure~\ref{fig:contour_enhancement_Tinf243_alpha8} shows how the ice enhancement is distributed
in the wake of a hydrometeor at $T_{\infty} = -30 \si{\celsius}$ for an exponent of $\alpha=8$. At this parameter point, which can be
considered as rather favorable, we observe that $f_i$ can locally reach extremely high values of the order of $10^4$ in the near vicinity of the
hydrometeor and significant values greater than $10$ can still be observed in the far wake. 

\begin{figure}[th]
  \centering
  \includegraphics[]{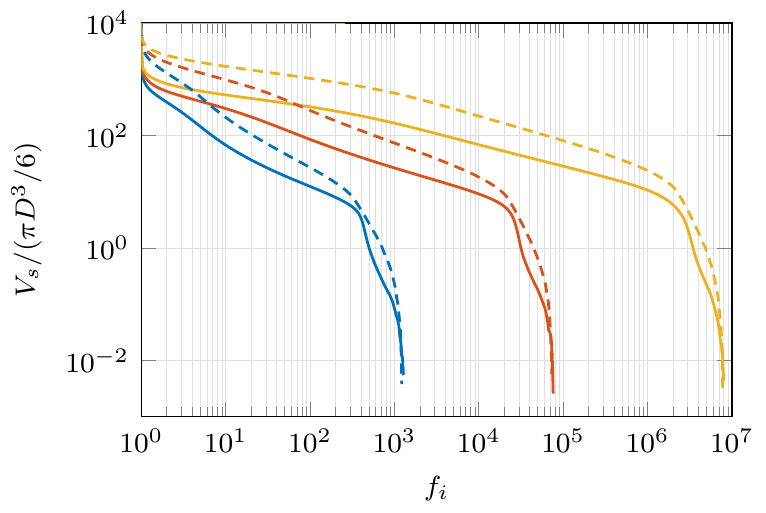}
  \caption{
      Volume of air with supersaturation above a given threshold as a function of the
      ice enhancement factor at $\alpha = 8$. The volume is normalized by the volume of the ice particle and three different temperatures
      are shown: 
      $T_{\infty} = -20\si{\celsius}$ ($\sampleline{mlb1}$),
      $T_{\infty} = -30\si{\celsius}$ ($\sampleline{mlb2}$),
      $T_{\infty} = -40\si{\celsius}$ ($\sampleline{mlb3}$).
      Solid lines correspond to $Re = 600$ (chaotic regime), while dashed lines show the data obtained for
      $Re = 75$ (axisymmetric regime).
  }
  \protect\label{fig:ice_enhancement_volume_cdf}
\end{figure}

Using Eq.~\eqref{eq:fix}, the supersaturated volume can be expressed as a function of the ice enhancement factor in order to gain insight into
the volumetric distribution of $f_i$. Figure~\ref{fig:ice_enhancement_volume_cdf} shows the distributions for $\alpha = 8$ and various 
temperatures in both wake regimes. The ice enhancement is generally higher in the axisymmetric regime than in the chaotic regime. 
Please note, that for a more precise analysis of this behavior, the actual flow-driven distribution of aerosol particles in the wake should be considered.
Depending on the flow dynamics, ice nucleating particles might accumulate in the wake or be expelled from it \citep{homann_concentrations_2015}, 
which might have a non-negligible impact on ice enhancement. However, the consideration of aerosol dynamics is outside the scope of
the present work.

We therefore conclude that the wake of a 
hydrometeor may act as a site of increased nucleation activity, and that the volume in which this increased activity occurs is much larger
than the volume of the hydrometeor itself. This conclusion is consistent with recent experimental results of \cite{prabhakaran_ice_2019}
who observed nucleation of water droplets and ice particles in the wake of a hot drop under cold conditions with a temperature difference
comparable to the one presented in fig.~\ref{fig:contour_enhancement_Tinf243_alpha8}.

In order to assess the significance of the wake effect in clouds, a global ice enhancement factor needs to be computed, which takes into
account the spatial distribution of $f_i$ obtained by the numerical simulations, as well as the volumetric ice concentration. 
We define a control volume $\mathcal{V}$ within 
a cloud which is sufficiently large such that the hydrometeor size distribution follows Eq.~\eqref{eq:marshall_palmer}, 
but sufficiently small such that the
saturation field is uniform (with a value of $s_{i,\infty}$) if no hydrometeor is present. 
The volume-averaged ice enhancement within this control volume is given by 
\begin{equation}
\left<{f_i}\right>_{\mathcal{V}} = \frac{\int_\mathcal{V} \del{f_i(\vec{x})-1} \dif \mathcal{V}}{\int_\mathcal{V} \dif \mathcal{V}} + 1,
\label{eq:glob-enh-average}
\end{equation}
where the integral has been decomposed into an excess contribution, which evaluates to zero outside
of the wake, and a base contribution, which is equal to unity. At sufficiently small volume fractions, the wakes of individual
hydrometeors can be assumed to not interact, and hence, the integral in the enumerator of the first term in 
Eq.~\eqref{eq:glob-enh-average} can be expressed 
as a sum of contributions from independent hydrometeors, i.e.
\begin{equation}
\int_\mathcal{V} \del{f_i(\vec{x})-1} \dif \mathcal{V}
= 
\sum_{(k)}\int_{\Omega^{(k)}} \del{f_i(\vec{x})-1} \dif \Omega^{(k)},
\end{equation}
where $\Omega^{(k)} \subseteq \mathcal{V}$ denotes the simulation domain scaled by the size of particle $(k)$.
By defining the excess ice enhancement per meteor volume as 
\begin{equation}
  \tilde{f}_{i,\Omega^{(k)}} \equiv \frac{6}{\pi D^3} \int_{\Omega^{(k)}}  \del{f_i(\vec{x}) - 1} \dif \Omega^{(k)}
\end{equation}
and assuming a continuous particle size distribution following Eq.~\eqref{eq:marshall_palmer}, the volume-averaged
ice enhancement within $\mathcal{V}$ can be expressed as
\begin{equation}
  \left<f_i\right>_{\mathcal{V}} = 1 + \int_0^{\infty} \tilde{f}_{i,\Omega} \frac{\pi D^3}{6} N_0 \exp\del{-\lambda D} \dif D .
  \label{eq:fiv_complete}
\end{equation}
Since $\tilde{f}_{i,\Omega}$ is normalized by the volume of the ice particle, its value only depends on the flow field in non-dimensional
form, i.e. the diameter dependency only enters through the value of the Reynolds number. For reasons of simplicity, we assume
that $\tilde{f}_{i,\Omega}$ can be approximated for a given regime by the value determined for a single Reynolds number within that regime,
which allows us to rewrite Eq.~\eqref{eq:fiv_complete} in the simplified form
\begin{equation}
  \left<f_i\right>_{\mathcal{V}} = 1 + \sum_j \phi_{j} \tilde{f}_{i,\Omega_{j}},
\end{equation}
where $\phi_{j}$ is the volume fraction of meteors in regime $j$, as defined in Eq.~\eqref{eq:phij}, and $\tilde{f}_{i,\Omega_{j}}$ the value of
the excess ice enhancement evaluated for a single value of the Reynolds number in that regime. As has already been discussed and shown in
fig.~\ref{fig:regime_contribution}, the axisymmetric and chaotic regimes contribute the most to the volume fraction. We therefore
disregard the contribution of the two other regimes and adjust the threshold of regime transition accordingly.
\begin{equation}
  \left<f_i\right>_{\mathcal{V}} = 1 + \phi_{axi} \tilde{f}_{i,\Omega_{axi}} + \phi_{cha} \tilde{f}_{i,\Omega_{cha}}
\end{equation}
Both $\phi_j$ and $\tilde{f}_{i,\Omega}$ are functions of ambient temperature, and thus the global ice enhancement factor originating from 
meteor wakes can be expressed as a function of cloud temperature. While the former decreases exponentially with decreasing temperature,
the latter exhibits a strong increase leading to a counteracting effect. Furthermore, a large number of ice particles pertain to the axisymmetric 
regime for low temperatures, which is more favorable in terms of ice enhancement.

Please be aware that even though our aim is to quantify the wake-induced ice formation, the volume fraction of ice
will be regarded as constant. The reason for this is that the approach
used in this study does not allow us to derive the time scales of nucleation nor growth of newly created ice, and thus, the
time-dependent coupling with the size distribution is inaccessible.
Therefore, the following considerations merely apply to the initial state of a possible rapid glaciation process.

\begin{figure}[t!]
  \centering
  \includegraphics{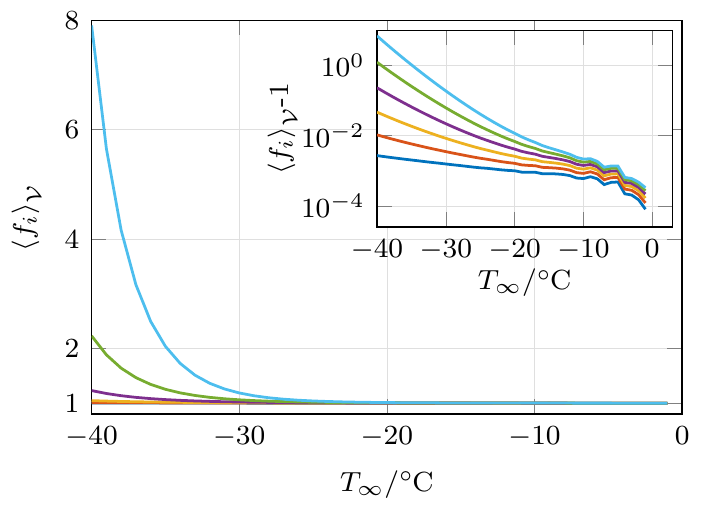}
  \caption{
      Global ice enhancement factor as a function of cloud temperature for various
      values of $\alpha$. The inset shows the same data, but in semi-logarithmic scale.
      Linestyles: $\alpha=3$ (\sampleline{mlb1}),
      $\alpha=4$ (\sampleline{mlb2}), $\alpha=5$ (\sampleline{mlb3}),
      $\alpha=6$ (\sampleline{mlb4}), $\alpha=7$ (\sampleline{mlb5}),
      $\alpha=8$ (\sampleline{mlb6}).
    }
  \protect\label{fig:ice_enhancement_global}
\end{figure}

Figure~\ref{fig:ice_enhancement_global} shows the global ice enhancement factor as a function of ambient temperature
for different values of $\alpha$, the exponent in Eq.~\eqref{eq:fix}. 
It can be seen that significant ice enhancement in the global sense only occurs
at cloud temperatures below $-30 \si{\celsius}$ in conjunction with high values of $\alpha$. 
Even then, the enhancement is only marginal with $\left<f_i\right>_{\mathcal{V}}$ being smaller than $8$. 
We therefore conclude that under typical cloud conditions hydrometeors do not directly lead to substantially enhanced
ice nucleation on a global scale.

Since the small volume fraction of ice is the limiting factor for ice enhancement on a global scale,
we revisited our assumption on the size distribution of ice particles in order to assess the sensitivity to the
parametrization. Using the parametrization of \cite{heymsfield_observations_2002} for deep subtropical and tropical clouds, 
the ice volume fraction may be up to five times larger at $T_{\infty} = -40 \si{\celsius}$. 
However, at higher cloud temperatures, i.e. when the temperature
difference between ice particles and the ambient is moderate, the volume fraction is of the same order of magnitude
as for the presented results, such that the conclusions which have been drawn for frontal clouds also persist for 
this type of cloud.

\section{Conclusion}
\label{sec:conclusion}
In this study we have performed numerical simulations of momentum, heat and 
mass transfer around a warm hydrometeor in order to assess the distribution
of supersaturation in its wake, and moreover, the implications on ice nucleation enhancement. 
Our simulation method is based on a body-conforming spectral/spectral-element discretization
and all relevant scales of the flow problem have been resolved. The hydrometeor is assumed to 
be of spherical shape and to possess a uniform surface temperature of $0\,\si{\celsius}$, while
the ambient temperature has been varied in a range between $-40\,\si{\celsius}$ and 
$0\,\si{\celsius}$. The vapor concentration is kept at saturation value with respect to ice 
at the particle surface, while it is saturated with respect to water in the ambient 
(reflecting the presence of supercooled droplets).
Two different values of the Reynolds number have been simulated in order to
capture the characteristics of the most relevant wake regimes, namely $Re = 75$, where
the wake is steady and axisymmetric, and $Re = 600$, where the wake is chaotic.

We found that significant values of supersaturation can be attained
in the wake of warm hydrometeors, which persist long enough to be observed at more than 50 particle
diameters downstream of the meteor for sufficiently high differences in temperature. The supersaturated 
volume of air exceeds the estimations by \cite{fukuta_numerical_1986} by far, which is attributed
to the more accurate representation of the flow in the current study. This is an important observation
since one of the key arguments for disregarding wake-induced ice nucleation is the proclaimed small 
zone of influence \citep{baker_nucleation_1991}.

Using a power-law approximation for the ice enhancement factor, we estimate that heterogeneous nucleation
may be locally enhanced by a factor comparable to the discrepancy between ice nuclei 
and ice particles observed in field measurements. This enhancement increases strongly with decreasing cloud 
temperature and so does the affected volume of air, which is typically of the order of several hundred 
particle diameters. However, due to the low volumetric concentration of ice in clouds, this local effect 
alone unlikely serves as an explanation for the global discrepancy, as arguments on upscaling have shown. 

Nonetheless, it is conceivable that the present mechanism in conjunction with one (or several) secondary
ice processes is of greater relevance to the problem of ice formation. After a rapid glaciation process 
has been triggered, wake-induced nucleation might become significant, as the ice concentration then increases 
to considerable values. 
Futhermore, the question of whether or not wake-induced nucleation alone might trigger such a process has
not been fully resolved yet, since the feedback on the size distribution of ice particles has been
disregarded in this study. In order to assess this dynamical behavior, further time-resolved information
on the (pre-)activation of ice nuclei in the wake currently appears to be necessary. 
This requires microphysical modeling
of the nucleation process and knowledge on the spatial distribution of aerosol particles downstream
of the ice particle. A resulting parametrization may be added to existing cloud models with 
explicit microphysics in order to assess the evolution of the size distribution, and moreover, the relative
significance of wake-induced nucleation in comparison to other secondary ice processes.

\paragraph{Data availability.}
The datasets are available upon request to the corresponding author.
\paragraph{Author contributions.}
TL and MU conceptualized the idea.  
JD provided the simulation code and mesh.
MK conducted the simulations.
MK, AC and MU post-processed the data and TL assisted in the interpretation of the results.
MK prepared the manuscript and AC contributed to it.
MU, AC, JD and TL revised the manuscript.
\paragraph{Competing interests.}
The authors declare that they have no conflict of interest.

\paragraph{Acknowledgements}
The authors thank Alexei Kiselev for useful discussions. The simulations were partially performed at 
the Steinbuch Centre for Computing in Karlsruhe and 
the computer resources, technical expertise and assistance provided by this center are thankfully acknowledged.

\bibliographystyle{apalike}
\bibliography{zotero}

\begin{thebibliography}{}

\bibitem[Auer et~al., 1969]{auer_observations_1969}
Auer, A.~H., Veal, D.~L., and Marwitz, J.~D. (1969).
\newblock Observations of {{Ice Crystal}} and {{Ice Nuclei Concentrations}} in
  {{Stable Cap Clouds}}.
\newblock {\em Journal of the Atmospheric Sciences}, 26(6):1342--1343.

\bibitem[Bacon et~al., 1998]{bacon_breakup_1998}
Bacon, N.~J., Swanson, B.~D., Baker, M.~B., and Davis, E.~J. (1998).
\newblock Breakup of levitated frost particles.
\newblock {\em Journal of Geophysical Research: Atmospheres},
  103(D12):13763--13775.

\bibitem[Bagchi et~al., 2001]{bagchi_direct_2001}
Bagchi, P., Ha, M.~Y., and Balachandar, S. (2001).
\newblock Direct {{Numerical Simulation}} of {{Flow}} and {{Heat Transfer
  From}} a {{Sphere}} in a {{Uniform Cross}}-{{Flow}}.
\newblock {\em Journal of Fluids Engineering}, 123(2):347--358.

\bibitem[Baker, 1991]{baker_nucleation_1991}
Baker, B.~A. (1991).
\newblock On the {{Nucleation}} of {{Ice}} in {{Highly Supersaturated Regions}}
  of {{Clouds}}.
\newblock {\em Journal of the Atmospheric Sciences}, 48(16):1904--1907.

\bibitem[Bouchet et~al., 2006]{bouchet_hydrodynamic_2006}
Bouchet, G., Mebarek, M., and Du{\v s}ek, J. (2006).
\newblock Hydrodynamic forces acting on a rigid fixed sphere in early
  transitional regimes.
\newblock {\em European Journal of Mechanics - B/Fluids}, 25(3):321--336.

\bibitem[Chouippe et~al., 2019]{chouippe_heat_2019}
Chouippe, A., Krayer, M., Uhlmann, M., Du{\v s}ek, J., Kiselev, A., and
  Leisner, T. (2019).
\newblock Heat and water vapor transfer in the wake of a falling ice sphere and
  its implication for secondary ice formation in clouds.
\newblock {\em New Journal of Physics}, 21(4):043043.

\bibitem[{de Stadler} et~al., 2014]{de_stadler_large_2014}
{de Stadler}, M.~B., Rapaka, N.~R., and Sarkar, S. (2014).
\newblock Large eddy simulation of the near to intermediate wake of a heated
  sphere at {{Re}}=10,000.
\newblock {\em International Journal of Heat and Fluid Flow}, 49:2--10.

\bibitem[Dye and Hobbs, 1968]{dye_influence_1968}
Dye, J.~E. and Hobbs, P.~V. (1968).
\newblock The {{Influence}} of {{Environmental Parameters}} on the {{Freezing}}
  and {{Fragmentation}} of {{Suspended Water Drops}}.
\newblock {\em Journal of the Atmospheric Sciences}, 25(1):82--96.

\bibitem[Field et~al., 2016]{field_secondary_2016}
Field, P.~R., Lawson, R.~P., Brown, P. R.~A., Lloyd, G., Westbrook, C.,
  Moisseev, D., Miltenberger, A., Nenes, A., Blyth, A., Choularton, T.,
  Connolly, P., Buehl, J., Crosier, J., Cui, Z., Dearden, C., DeMott, P.,
  Flossmann, A., Heymsfield, A., Huang, Y., Kalesse, H., Kanji, Z.~A., Korolev,
  A., Kirchgaessner, A., {Lasher-Trapp}, S., Leisner, T., McFarquhar, G.,
  Phillips, V., Stith, J., and Sullivan, S. (2016).
\newblock Secondary {{Ice Production}}: {{Current State}} of the {{Science}}
  and {{Recommendations}} for the {{Future}}.
\newblock {\em Meteorological Monographs}, 58:7.1--7.20.

\bibitem[Fukuta and Lee, 1986]{fukuta_numerical_1986}
Fukuta, N. and Lee, H.~J. (1986).
\newblock A {{Numerical Study}} of the {{Supersaturation Field}} around
  {{Growing Graupel}}.
\newblock {\em Journal of the Atmospheric Sciences}, 43(17):1833--1843.

\bibitem[Gagin, 1972]{gagin_effect_1972}
Gagin, A. (1972).
\newblock The effect of supersaturation on the ice crystal production by
  natural aerosols.
\newblock {\em Journal de Recherches Atmosph{\'e}riques}, 6:175--185.

\bibitem[Ghidersa and Du{\v s}ek, 2000]{ghidersa_breaking_2000}
Ghidersa, B. and Du{\v s}ek, J. (2000).
\newblock Breaking of axisymmetry and onset of unsteadiness in the wake of a
  sphere.
\newblock {\em Journal of Fluid Mechanics}, 423:33--69.

\bibitem[Gordon and Marwitz, 1984]{gordon_airborne_1984}
Gordon, G.~L. and Marwitz, J.~D. (1984).
\newblock An {{Airborne Comparison}} of {{Three PMS Probes}}.
\newblock {\em Journal of Atmospheric and Oceanic Technology}, 1(1):22--27.

\bibitem[Hallett and Mossop, 1974]{hallett_production_1974}
Hallett, J. and Mossop, S.~C. (1974).
\newblock Production of secondary ice particles during the riming process.
\newblock {\em Nature}, 249(5452):26--28.

\bibitem[Herzegh and Hobbs, 1985]{herzegh_size_1985}
Herzegh, P.~H. and Hobbs, P.~V. (1985).
\newblock Size spectra of ice particles in frontal clouds: {{Correlations}}
  between spectrum shape and cloud conditions.
\newblock {\em Quarterly Journal of the Royal Meteorological Society},
  111(468):463--477.

\bibitem[Heymsfield et~al., 2002]{heymsfield_observations_2002}
Heymsfield, A.~J., Bansemer, A., Field, P.~R., Durden, S.~L., Stith, J.~L.,
  Dye, J.~E., Hall, W., and Grainger, C.~A. (2002).
\newblock Observations and {{Parameterizations}} of {{Particle Size
  Distributions}} in {{Deep Tropical Cirrus}} and {{Stratiform Precipitating
  Clouds}}: {{Results}} from {{In Situ Observations}} in {{TRMM Field
  Campaigns}}.
\newblock {\em Journal of the Atmospheric Sciences}, 59(24):3457--3491.

\bibitem[Hobbs, 1969]{hobbs_ice_1969}
Hobbs, P.~V. (1969).
\newblock Ice {{Multiplication}} in {{Clouds}}.
\newblock {\em Journal of the Atmospheric Sciences}, 26(2):315--318.

\bibitem[Hobbs and Alkezweeny, 1968]{hobbs_fragmentation_1968}
Hobbs, P.~V. and Alkezweeny, A.~J. (1968).
\newblock The {{Fragmentation}} of {{Freezing Water Droplets}} in {{Free
  Fall}}.
\newblock {\em Journal of the Atmospheric Sciences}, 25(5):881--888.

\bibitem[Hobbs and Rangno, 1985]{hobbs_ice_1985}
Hobbs, P.~V. and Rangno, A.~L. (1985).
\newblock Ice {{Particle Concentrations}} in {{Clouds}}.
\newblock {\em Journal of the Atmospheric Sciences}, 42(23):2523--2549.

\bibitem[Hogan et~al., 2002]{hogan_properties_2002}
Hogan, R.~J., Field, P.~R., Illingworth, A.~J., Cotton, R.~J., and Choularton,
  T.~W. (2002).
\newblock Properties of embedded convection in warm-frontal mixed-phase cloud
  from aircraft and polarimetric radar.
\newblock {\em Quarterly Journal of the Royal Meteorological Society},
  128(580):451--476.

\bibitem[Homann and Bec, 2015]{homann_concentrations_2015}
Homann, H. and Bec, J. (2015).
\newblock Concentrations of inertial particles in the turbulent wake of an
  immobile sphere.
\newblock {\em Physics of Fluids}, 27(5):053301.

\bibitem[Houze et~al., 1979]{houze_size_1979}
Houze, R.~A., Hobbs, P.~V., Herzegh, P.~H., and Parsons, D.~B. (1979).
\newblock Size {{Distributions}} of {{Precipitation Particles}} in {{Frontal
  Clouds}}.
\newblock {\em Journal of the Atmospheric Sciences}, 36(1):156--162.

\bibitem[Huffman, 1973]{huffman_supersaturation_1973}
Huffman, P.~J. (1973).
\newblock Supersaturation {{Spectra}} of {{AgI}} and {{Natural Ice Nuclei}}.
\newblock {\em Journal of Applied Meteorology (1962-1982)}, 12(6):1080--1082.

\bibitem[Jenny and Du{\v s}ek, 2004]{jenny_efficient_2004}
Jenny, M. and Du{\v s}ek, J. (2004).
\newblock Efficient numerical method for the direct numerical simulation of the
  flow past a single light moving spherical body in transitional regimes.
\newblock {\em Journal of Computational Physics}, 194(1):215--232.

\bibitem[Jenny et~al., 2004]{jenny_instabilities_2004}
Jenny, M., Du{\v s}ek, J., and Bouchet, G. (2004).
\newblock Instabilities and transition of a sphere falling or ascending freely
  in a {{Newtonian}} fluid.
\newblock {\em Journal of Fluid Mechanics}, 508:201--239.

\bibitem[Johnson and Patel, 1999]{johnson_flow_1999}
Johnson, T.~A. and Patel, V.~C. (1999).
\newblock Flow past a sphere up to a {{Reynolds}} number of 300.
\newblock {\em Journal of Fluid Mechanics}, 378:19--70.

\bibitem[Koenig, 1963]{koenig_glaciating_1963}
Koenig, L.~R. (1963).
\newblock The {{Glaciating Behavior}} of {{Small Cumulonimbus Clouds}}.
\newblock {\em Journal of the Atmospheric Sciences}, 20(1):29--47.

\bibitem[Korolev et~al., 2020]{korolev_new_2020}
Korolev, A., Heckman, I., Wolde, M., Ackerman, A.~S., Fridlind, A.~M., Ladino,
  L.~A., Lawson, R.~P., Milbrandt, J., and Williams, E. (2020).
\newblock A new look at the environmental conditions favorable to secondary ice
  production.
\newblock {\em Atmospheric Chemistry and Physics}, 20(3):1391--1429.

\bibitem[Kotou{\v c} et~al., 2008]{kotouc_loss_2008}
Kotou{\v c}, M., Bouchet, G., and Du{\v s}ek, J. (2008).
\newblock Loss of axisymmetry in the mixed convection, assisting flow past a
  heated sphere.
\newblock {\em International Journal of Heat and Mass Transfer},
  51(11-12):2686--2700.

\bibitem[Kotou{\v c} et~al., 2009]{kotouc_transition_2009}
Kotou{\v c}, M., Bouchet, G., and Du{\v s}ek, J. (2009).
\newblock Transition to turbulence in the wake of a fixed sphere in mixed
  convection.
\newblock {\em Journal of Fluid Mechanics}, 625:205.

\bibitem[Marshall and Palmer, 1948]{marshall_distribution_1948}
Marshall, J.~S. and Palmer, W. M.~K. (1948).
\newblock The distribution of raindrops with size.
\newblock {\em Journal of Meteorology}, 5(4):165--166.

\bibitem[Mossop, 1985]{mossop_origin_1985}
Mossop, S.~C. (1985).
\newblock The {{Origin}} and {{Concentration}} of {{Ice Crystals}} in
  {{Clouds}}.
\newblock {\em Bulletin of the American Meteorological Society},
  66(3):264--273.

\bibitem[Murphy and Koop, 2005]{murphy_review_2005}
Murphy, D.~M. and Koop, T. (2005).
\newblock Review of the vapour pressures of ice and supercooled water for
  atmospheric applications.
\newblock {\em Quarterly Journal of the Royal Meteorological Society},
  131(608):1539--1565.

\bibitem[Nix and Fukuta, 1974]{nix_nonsteady-state_1974}
Nix, N. and Fukuta, N. (1974).
\newblock Nonsteady-{{State Kinetics}} of {{Droplet Growth}} in {{Cloud
  Physics}}.
\newblock {\em Journal of the Atmospheric Sciences}, 31(5):1334--1343.

\bibitem[Ormi{\`e}res et~al., 1999]{ormieres_transition_1999}
Ormi{\`e}res, D., Provansal, M., Recherche, I. R. P. H. E. I.~D., Hors, P., and
  Cnrs, U. M.~R. (1999).
\newblock Transition to {{Turbulence}} in the {{Wake}} of a {{Sphere}}.
\newblock {\em Physical Review Letters}, pages 6--9.

\bibitem[Passarelli, 1978]{passarelli_theoretical_1978}
Passarelli, R.~E. (1978).
\newblock Theoretical and {{Observational Study}} of {{Snow}}-{{Size Spectra}}
  and {{Snowflake Aggregation Efficiencies}}.
\newblock {\em Journal of the Atmospheric Sciences}, 35(5):882--889.

\bibitem[Patade et~al., 2015]{patade_particle_2015}
Patade, S., Prabha, T.~V., Axisa, D., Gayatri, K., and Heymsfield, A. (2015).
\newblock Particle size distribution properties in mixed-phase monsoon clouds
  from in situ measurements during {{CAIPEEX}}.
\newblock {\em Journal of Geophysical Research: Atmospheres},
  120(19):10,418--10,440.

\bibitem[Patera, 1984]{patera_spectral_1984}
Patera, A.~T. (1984).
\newblock A spectral element method for fluid dynamics: {{Laminar}} flow in a
  channel expansion.
\newblock {\em Journal of Computational Physics}, 54(3):468--488.

\bibitem[Prabhakaran et~al., 2019]{prabhakaran_ice_2019}
Prabhakaran, P., Kinney, G., Cantrell, W., Shaw, R.~A., and Bodenschatz, E.
  (2019).
\newblock Ice nucleation in the wake of warm hydrometeors.
\newblock {\em arXiv:1906.06129 [physics]}.

\bibitem[Prabhakaran et~al., 2017]{prabhakaran_can_2017}
Prabhakaran, P., Weiss, S., Krekhov, A., Pumir, A., and Bodenschatz, E. (2017).
\newblock Can {{Hail}} and {{Rain Nucleate Cloud Droplets}}?
\newblock {\em Physical Review Letters}, 119(12):128701.

\bibitem[Pruppacher and Klett, 2010]{pruppacher_microphysics_2010}
Pruppacher, H.~R. and Klett, J.~D. (2010).
\newblock {\em Microphysics of {{Clouds}} and {{Precipitation}}}.
\newblock {Springer Netherlands}.

\bibitem[Rangno and Hobbs, 1991]{rangno_ice_1991}
Rangno, A.~L. and Hobbs, P.~V. (1991).
\newblock Ice particle concentrations and precipitation development in small
  polar maritime cumuliform clouds.
\newblock {\em Quarterly Journal of the Royal Meteorological Society},
  117(497):207--241.

\bibitem[R{\o}nquist, 1988]{ronquist_optimal_1988}
R{\o}nquist, E.~M. (1988).
\newblock {\em Optimal Spectral Element Methods for the Unsteady
  Three-Dimensional Incompressible {{Navier}}-{{Stokes}} Equations}.
\newblock PhD thesis, Massachusetts Institute of Technology.

\bibitem[Schiller and Naumann, 1933]{schiller_uber_1933}
Schiller, L. and Naumann, A. (1933).
\newblock {\"U}ber die grundlegenden {{Berechnungen}} bei der
  {{Schwerkraftaufbereitung}}.
\newblock {\em Z. Ver. Dtsch. Ing}, 77(12):318--320.

\bibitem[Takahashi et~al., 1995]{takahashi_possible_1995}
Takahashi, T., Nagao, Y., and Kushiyama, Y. (1995).
\newblock Possible {{High Ice Particle Production}} during
  {{Graupel}}\textendash{{Graupel Collisions}}.
\newblock {\em Journal of the Atmospheric Sciences}, 52(24):4523--4527.

\bibitem[Vardiman, 1978]{vardiman_generation_1978}
Vardiman, L. (1978).
\newblock The {{Generation}} of {{Secondary Ice Particles}} in {{Clouds}} by
  {{Crystal}}\textendash{{Crystal Collision}}.
\newblock {\em Journal of the Atmospheric Sciences}, 35(11):2168--2180.

\bibitem[Zhou and Du{\v s}ek, 2015]{zhou_chaotic_2015}
Zhou, W. and Du{\v s}ek, J. (2015).
\newblock Chaotic states and order in the chaos of the paths of freely falling
  and ascending spheres.
\newblock {\em International Journal of Multiphase Flow}, 75:205--223.

\end{thebibliography}

\end{document}